\documentclass[conference]{IEEEtran}

\usepackage{booktabs} % For formal tables
\usepackage[pdftex]{graphicx}
\begin{document}
\title{Serving deep learning models in a serverless platform}
%\title{On the feasibility of serving deep learning models on a serverless platform}
%\title{Serverless and AI: match made in heaven or hell? Current state and future steps}
%\titlenote{Produces the permission block, and copyright information}
%\subtitle{Extended Abstract}
%\subtitlenote{The full version of the author's guide is available as   \texttt{acmart.pdf} document}

% \author{Vatche Ishakian}
% %\authornote{Dr.~Trovato insisted his name be first.}
% \affiliation{%
%   \institution{Bentley University}
%   \city{Waltham} 
%   \state{MA} 
% }
% \email{vishakian@bentley.edu}

% \author{Vinod Muthusamy}
% \affiliation{%
%   \institution{IBM T.J. Watson Research Center}
%   \city{Yorktown Heights} 
%   \state{NY}
% }
% \email{vmuthus@us.ibm.com}

% \author{Aleksander Slominski}
% \affiliation{%
%   \institution{IBM T.J. Watson Research Center}
%   \city{Yorktown Heights} 
%   \state{NY}
% }
% \email{aslom@us.ibm.com}

\author{\IEEEauthorblockN{Vatche Ishakian}
\IEEEauthorblockA{vishakian@bentley.edu}
\IEEEauthorblockA{Bentley University}
\and
\IEEEauthorblockN{Vinod Muthusamy}
\IEEEauthorblockA{vmuthus@us.ibm.com}
\IEEEauthorblockA{IBM T.J. Watson Research Center}
\and
\IEEEauthorblockN{Aleksander Slominski}
\IEEEauthorblockA{aslom@us.ibm.com}
\IEEEauthorblockA{IBM T.J. Watson Research Center}
}

% \begin{abstract}
% Serverless computing has emerged as a compelling paradigm for the development and deployment of a wide range of event based cloud applications. At the same time, cloud providers and enterprise companies are heavily adopting machine learning and Artificial Intelligence to either differentiate themselves, or provide their customers with value added services. 
% In this work we evaluate the suitability of a serverless computing environment for the inferencing of large neural network models. Our experimental evaluations are executed on the AWS Lambda environment using the MxNet deep learning framework.
% Our experimental results show that while the inferencing latency can be within an acceptable range, longer delays due to cold starts can skew the latency distribution and hence risk violating more stringent SLAs.
% \end{abstract}


% \ccsdesc[500]{Software and its engineering}
% \ccsdesc[300]{Distributed systems organizing principles~Cloud computing}
% \ccsdesc{Computer systems organization~Cloud computing}

% \keywords{Serverless Computing, Function as a Service,  Cloud Computing, Cognitive Computing, Machine Learning, Deep Learning, AI,}

\maketitle
\begin{abstract}
Serverless computing has emerged as a compelling paradigm for the development and deployment of a wide range of event based cloud applications. At the same time, cloud providers and enterprise companies are heavily adopting machine learning and Artificial Intelligence to either differentiate themselves, or provide their customers with value added services. 
In this work we evaluate the suitability of a serverless computing environment for the inferencing of large neural network models. Our experimental evaluations are executed on the AWS Lambda environment using the MxNet deep learning framework.
Our experimental results show that while the inferencing latency can be within an acceptable range, longer delays due to cold starts can skew the latency distribution and hence risk violating more stringent SLAs.
\end{abstract}
\section{Introduction}

Serverless computing, and in particular functions-as-a-service (FaaS), has emerged as a compelling paradigm for the deployment of cloud applications, largely due to the recent shift of enterprise application architectures to containers and microservices~\cite{fox2017status,baldini2017serverless}.
Serverless platforms promise new capabilities that make writing scalable microservices easier and cost effective, positioning themselves as the next step in the evolution of cloud computing architectures. It has been utilized to support a wider range of applications \cite{baldini2017serverless,Chard2017WOSC,yan2016building,baldini2016cloud,bila2017leveraging}. 
%Most of the prominent cloud providers including Amazon, IBM, Microsoft, and Google offer serverless computing capabilities.

From the perspective of a cloud customer, serverless platforms provide developers with a simplified programming model for their applications that abstracts away most, if not all, operational concerns; lowers cost by charging for execution time rather than resource allocation; and enables rapidly deploying small pieces of cloud-native event handlers.
For a cloud provider, serverless computing provides an opportunity to control the entire development stack, reduce operational costs by efficient optimization and management of cloud resources, and encourage the use of an ecosystem of additional cloud provider services. 

Recently, there has been a surge in the use of machine learning and artificial intelligence (AI) by both cloud providers and enterprise companies as value added differentiating services. In particular, we refer to the use of deep learning, a particular field of machine learning that utilizes neural networks to provide services that range from text analytics, natural language processing, and speech and image recognition.
Training neural networks is a time consuming job that typically requires specialized hardware, such as GPUs. Using a trained neural network for inferencing (a.k.a serving), on the other hand, requires no training data and less computational power than training them. 

In this work, we attempt to evaluate the suitability of serverless computing to run deep learning inferencing tasks. We measure the performance as seen by the user, and the cost of running three different MXNet \cite{chen2015mxnet} trained deep learning models on the AWS Lambda serverless computing platform. Our preliminary set of experimental results show that a serverless platform is suitable for deploying deep learning prediction services, but the delays occured as a result of running the first lambda invocation can skew the latency distribution and hence risk violating more stringent SLAs.

%We aim to understand the tradeoffs between different dimensions of serverless computing, mainly cost, memory and total latency, to support the next generation of services. We also evaluate whether Serverless computing is suitable to run applications that are not 

The remainder of this paper is organized as follows. Section~\ref{sec:background} introduces serverless computing and deep learning concepts. Section~\ref{sec:expsetup} presents our experimental setup and performance evaluations, with a discussion of findings and limitations. Section~\ref{sec:relatedwork} reviews related work, and Section~\ref{sec:conclusion} presents closing remarks and future research directions.
 
\section{Background} \label{sec:background}
\subsection{Serverless Computing}
The core capability of any serverless computing platform, as exemplified by the Apache OpenWhisk platform~\cite{apacheopenwhisk}, is that of an event processing system that utilizes container technology\footnote{Other serverless platforms such as AWS Lambda~\cite{awslambdacontainers} also rely on container technology to provide their serverless capabilities}. The service must manage a set of user defined functions, take an event received from an event source (e.g., http) determine which function to dispatch the event to, launch one or more containers to execute the function, send the event to the function instance, wait for a response, make the response and execution logs available to the client, and stop the function when it is no longer needed.

Setting up a container and doing the necessary bootstrapping typically takes some time, which adds  latency whenever a serverless function is invoked. This additional latency, which is referred to as the cold start phenomenon, is typically observed by the client when the serverless function is invoked for the first time. To minimize that latency and avoid bootstrapping time, the platform tries to reuse the container for subsequent invocations of the serverless function. The latency observed by the client when a container is reused is called a warm start.

\subsection{Deep Learning}
Deep learning, driven by large neural network models, is overtaking traditional machine learning methods for understanding unstructured and perceptual data domains such as speech, text, and vision. The rise of deep learning \cite{lecun2015deep} from its roots to becoming the state-of-the-art of AI has been fueled by three recent trends: the explosion in the amount of training data, the use of accelerators such as graphics processing units (GPUs), and advancements in the design of models used for training. These three trends have made the task of training deep layer neural networks with large amounts of data both tractable and useful.

Using any of the deep learning frameworks (e.g., Caffe \cite{jia2014caffe}, Tensorflow \cite{abadi2016tensorflow}, MXNet \cite{chen2015mxnet}), users can develop and train their models. Neural network models range in size from small (5MB) to very large (500MB). Training neural networks can take a significant amount of time, and the goal is to find suitable weights for the different variables in the neural network. 
Once the model training is complete, it can be used for inferencing --serving -- : applying the trained model on new data in domains such a natural language processing, speech recognition, or image classification.

\section{Experiments} \label{sec:expsetup}
As we noted above, one of the goals of our evaluation is to understand if a serverless computing platform can be utilized for neural network inferencing. 

The experiments use the AWS Lambda serverless platform, and the Amazon MXNet deep learning framework.
%The reason we chose AWS lambda and MXNet is mostly because they are both Amazon AWS products hence it was relatively easy to integrate them, and the publicly available trained models~\cite{lambdamodels}.
In the future, we plan to extend our evaluations to include other serverless platforms and deep learning frameworks, such as Tensorflow among others.

The model size is an important factor that affects the inferencing performance. We evaluate three popular image recognition models that represent a spectrum of model sizes:
\begin{itemize}
\item
SqueezeNet \cite{SqueezeNet} is among the smallest models available with reasonably high accuracy. We chose SqueezeNet V1.0 which has a relatively small size of 5 MB.  %The model provides an Alex-Net level accuracy in prediction with 50x fewer parameter. 
\item
ResNet \cite{Kaiming2015} is an easy to optimize deep residual learning model. ResNet models can vary in size depending on the complexity of layers used in the models. Our choice of ResNet-18 consists of 18 layers and has a size of 45 MB.
\item 
ResNeXt-50 \cite{Xie2016} is a simple, highly modularized network architecture, constructed by repeating a building block that aggregates a set of transformations with the same topology. ResNeXt models can have varying number of layers. Our choice of ResNeXt-50 has a total size of 98MB.
\end{itemize}

We adapted publicly available code from  AWSLabs~\cite{awslabscode}, which downloads an MXNet model from S3 and also loads an image -- from a specified URL -- to classify by performing a forward pass through the model. To factor out delays that may occur as a result of model or image downloads, we included both the image as well as the model as part of AWS lambda function dependency libraries. We also modified the output to include only the total prediction time. We created a zip file with the dependency and used the AWS GUI interface to create Lambda functions. The maximum amount of memory  used by a SqueezeNet-Lambda prediction function during execution is 85MB, which compares to 229MB for a ResNet-Lambda prediction function, and 429MB for a ResNeXt-Lambda prediction function. We use Amazon API Gateway~\cite{apigateway} to provide a restful endpoint for our Lambda functions, making them accessible with an HTTP GET request. 

AWS Lambda allows its clients the choice between different memory sizes. The size of the memory ranges from 128MB to 1536 MB going up in increments of 64MB. The AWS Lambda  platform allocates other resources such as CPU power, network bandwidth and disk I/O in proportion to the choice of memory~\cite{awslambdafaq}. The cost of running a Lambda function is measured in 100 millisecond intervals. Table~\ref{tab:lambdaPrice} shows the cost for varying memory sizes\footnote{We report only the cost associated with the memory size that we used throughout our experiments.}.

\begin{table}[htp]
  \centering
  \begin{tabular}{|c|c|}
  \hline
    Memory (MB) & Price per 100ms (\$) \\
   \hline
  128 & 0.000000208 \\
  \hline
  256 & 0.000000417 \\
  \hline
  384 & 0.000000625 \\
  \hline
    512 & 0.000000834 \\
  \hline
    640 & 0.000001042 \\
  \hline
    768 & 0.00000125 \\
  \hline
    896 & 0.000001459 \\
  \hline
    1024 & 0.000001667 \\
  \hline
    1152 & 0.000001875 \\
  \hline
    1280 & 0.000002084 \\
  \hline
    1408 & 0.000002292 \\
  \hline
    1536 & 0.000002501 \\
  \hline
  \end{tabular}
  \caption{AWS Lambda price per 100ms of execution associated for different memory sizes.}
  \label{tab:lambdaPrice}
\end{table}

We used Apache JMeter \cite{jmeter}, a pure Java application originally designed to load test the functional behavior and measure performance of web applications, but has expanded to include other types of test functions to issue http get requests to our lambda functions.
%To measure the response time from the users point of view and enable scalability experiments. 

Our focus is on measurements under two particular situations in a serverless computing environment, which are labeled as cold start and warm start. A cold start refers to a situation where the container needs to be initialized and in a running state before the Lambda function is executed. Needless to say that this will incur an additional overhead on the system and increased delay as observed from the users point of view. A warm start refers to a situation where the container is initialized and is in a running state, hence the overhead of execution is that of the Lambda function itself.

We capture the following metrics:
\begin{itemize}
\item Response time: The latency as observed by the user both for warm and cold environments.
\item Prediction time: The total time it takes for the model to return with a prediction.
\item Cost: Total cost of executing the Lambda function.
\end{itemize}

\subsection{Cold and warm evaluations}
To measure cold starts we configure our JMeter script to send 5 sequential HTTP requests to the Lambda function separated by 10 minutes of wait time. To measure warm startup time, we configure our script to send a request, disregard it, then send 25 sequential requests to the Lambda function where each request is separated by one second intervals. All results are reported with 95\% confidence.

\subsection{Warm Results}
\begin{figure}[h]
\centering
\includegraphics[width=0.50\textwidth]{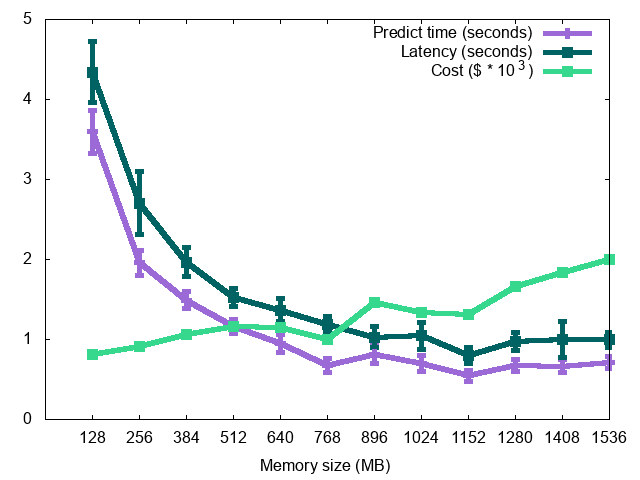}
\caption{Warm function execution (SqueezeNet)}
\label{squeezenet-warm}
\end{figure}

\begin{figure}[h]
\centering\includegraphics[width=0.50\textwidth]{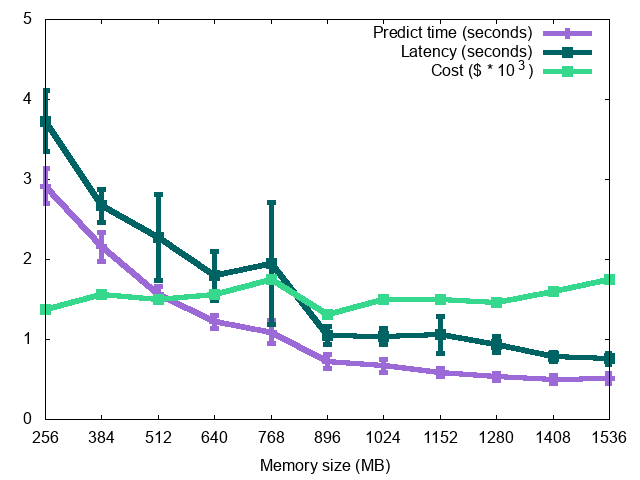}
\caption{Warm function execution (Resnet)}
\label{resnet-warm}
\end{figure}

\begin{figure}[h]
\centering
\includegraphics[width=0.50\textwidth]{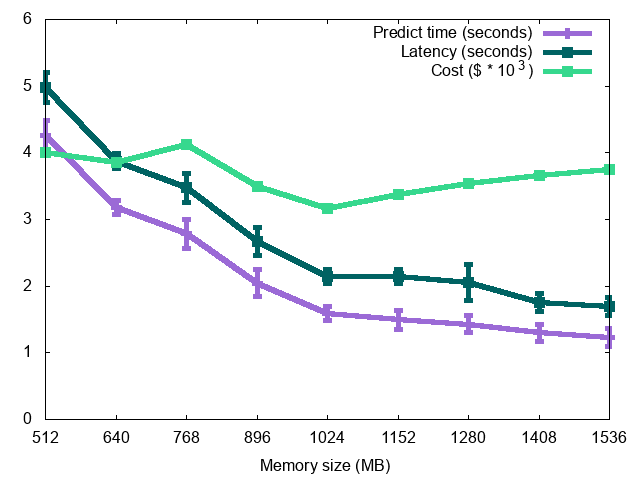}
\caption{Warm function execution (ResNeXt)}
\label{resnext-warm}
\end{figure}

Figure \ref{squeezenet-warm} shows the results of our warm experiments for a Lambda function that utilizes the SqueezeNet neural network model for prediction. The x-axis shows varying amount of Lambda function memory sizes. The figure shows the average latency in seconds as observed by the Jmeter client, the average prediction time in seconds, and the total cost of executing the function in dollars multiplied by one thousand\footnote{We multiplied the cost of execution by $10^3$ for better display.}.  Figures \ref{resnet-warm} and \ref{resnext-warm} show similar results for the ResNet and ResNeXt models respectively. 

The results show that total prediction time follows a similar pattern to the latency. That is expected since the the prediction time is a component in the total latency.

The results also show that as memory size increases the total delay and prediction time decreases. Initially, we suspect that this is due to the amount of physical memory allocated to run the Lambda function. That does not turn our to be the case, because when looking at the Lambda function execution logs, the maximum amount of memory measured to run the function is 85MB (and 229MB and 429MB for ResNet and ResNeXt, respectively). Since the platform assigns CPU, disk, and I/O resources in proportion to memory allocation, we believe that the delay incurred is due to varying CPU resources.

An interesting observation is that the total cost of executing the Lambda functions does not necessarily increase with the memory size. While the cost of each 100ms execution time unit increases with memory size ({\it c.f. } Table \ref{tab:lambdaPrice}), the decrease in total execution execution time offsets the cost.

The figures also show that an increase in memory size does not necessarily lead to improvements in performance. Particularly, Figure \ref{squeezenet-warm} shows that as we increase memory size from 1024MB to 1536MB there is no considerable improvement but the total cost of execution increases. Similar observations can be made for figures \ref{resnet-warm} and \ref{resnext-warm}. This could be problematic since a customer may incur additional costs of allocating more resources than what the function needs to execute. 

\begin{figure}[h]
\centering
\includegraphics[width=0.50\textwidth]{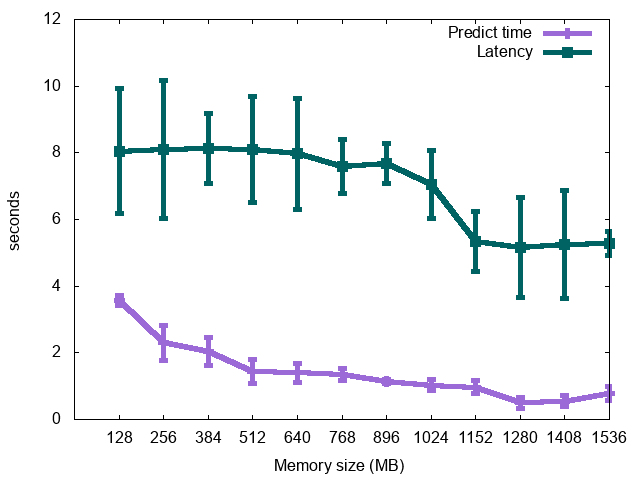}
\caption{Cold function execution (SqueezeNet)}
\label{squeezenet-cold}
\end{figure}

\begin{figure}[h]
\centering\includegraphics[width=0.50\textwidth]{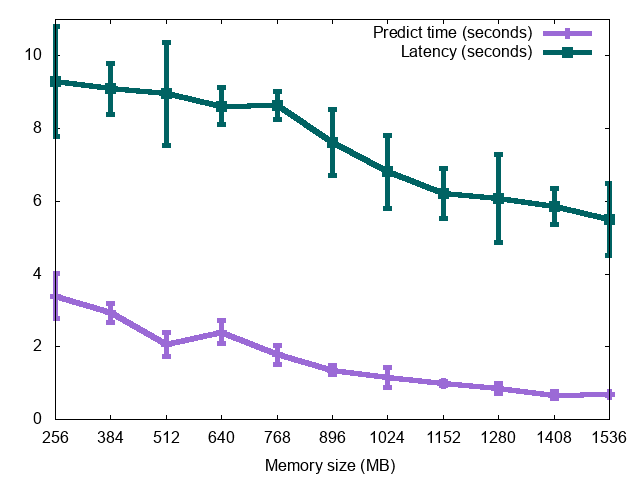}
\caption{Cold function execution (ResNet)}
\label{resnet-cold}
\end{figure}

\subsection{Cold Results}

\begin{figure}[h]
\centering
\includegraphics[width=0.50\textwidth]{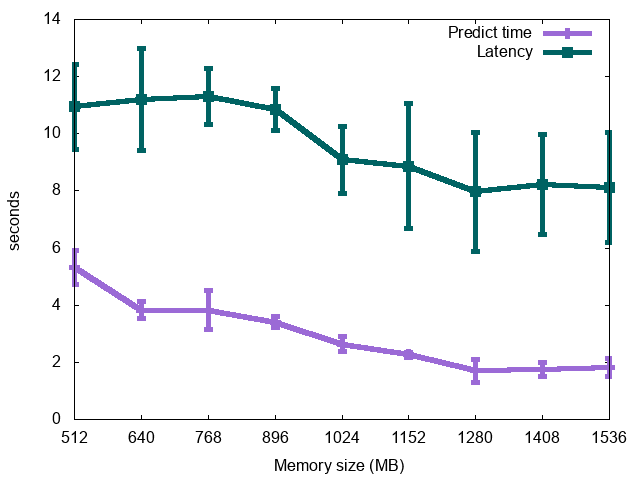}
\caption{Cold function execution (ResNeXt)}
\label{resnext-cold}
\end{figure}

Figure \ref{squeezenet-cold} shows the results of our cold experiments for a Lambda function that utilizes the SqueezeNet model for prediction. The x-axis shows a varying amount of Lambda function memory sizes and the y-axis shows the time in seconds. The Figure shows the average latency in seconds as observed by the JMeter client, and the average prediction time. Figures \ref{resnet-cold} and \ref{resnext-cold} show similar results for the ResNet and ResNeXt models, respectively. 

The figures show that while the cold start times decrease as we increase the memory size, it does not follow a similar pattern to that of warm starts. We believe this is because the overhead of launching and bootstrapping a container is the one that is dominating.

%VM: Why are there no cost numbers for the cold start experiments?

\begin{figure}[h]
\centering
\includegraphics[width=0.50\textwidth]{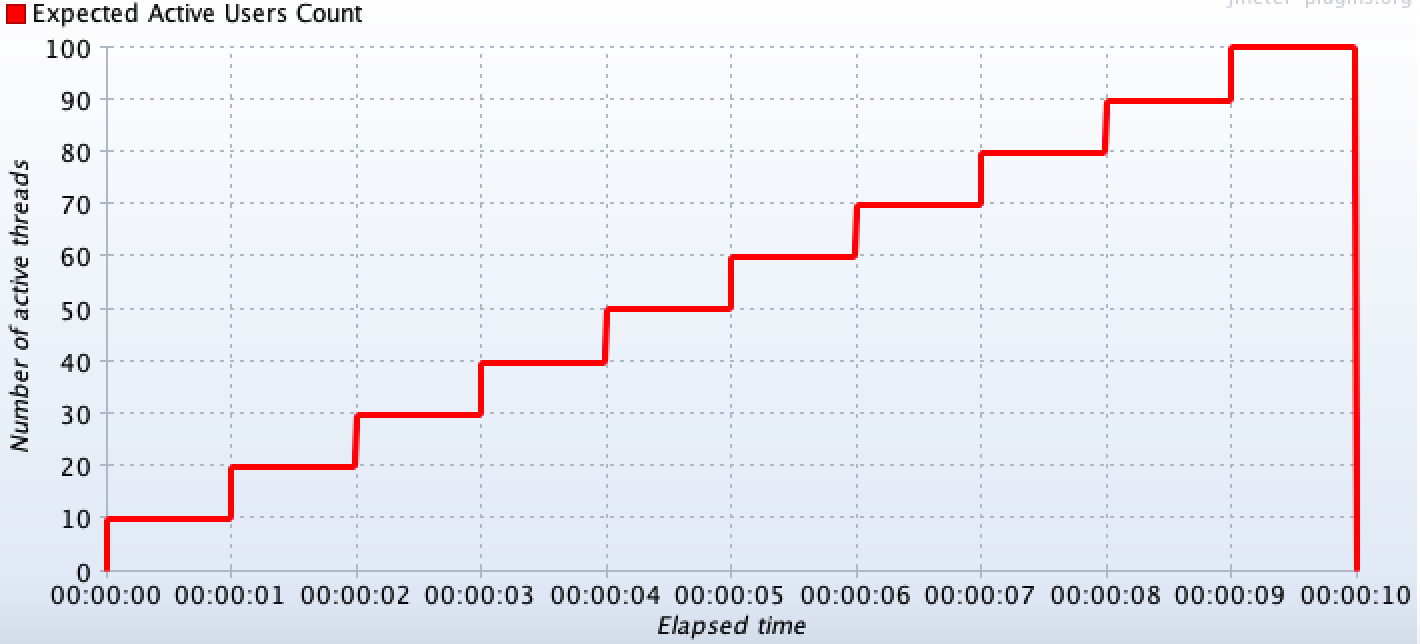}
\caption{Step function representing the increased number of lambda function requests.}
\label{jmeter:scalability}
\end{figure}

\subsection{Scalability evaluations}
Our scalability evaluations aim to measure the ability of AWS Lambda to seamlessly scale with the increase in demand. We configure our JMeter script to generate 10 HTTP requests in parallel and increase requests rates by 10 requests per second for 10 seconds. Figure \ref{jmeter:scalability} highlights the JMeter configuration used. Note that during this experiment, we are not able to distinguish between a warm and a cold start since we are not sure whether our request is routed to an existing container or a newly launched one.

\begin{figure}[h]
\centering
\includegraphics[width=0.50\textwidth]{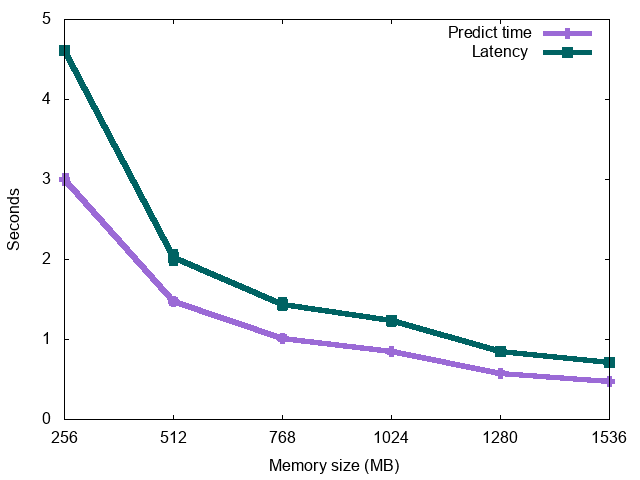}
\caption{Scalable lambda function execution results: SqueezeNet neural network}
\label{squeezenet-scalable}
\end{figure}

\begin{figure}[h]
\centering\includegraphics[width=0.50\textwidth]{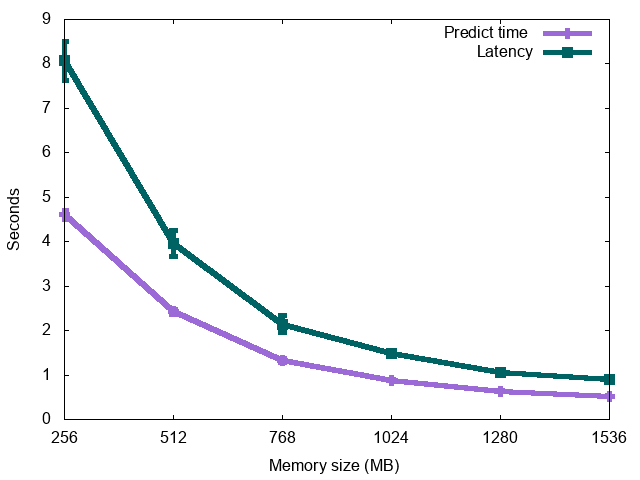}
\caption{Scalable lambda function execution results: resnet neural network}
\label{resnet-scalable}
\end{figure}

\begin{figure}[h]
\centering
\includegraphics[width=0.50\textwidth]{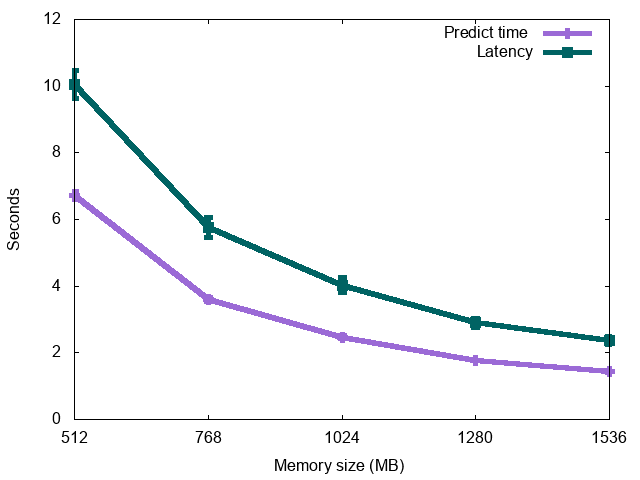}
\caption{Scalable lambda function execution results: resnext neural network}
\label{resnext-scalable}
\end{figure}

Figure \ref{squeezenet-scalable} shows the results of our experiments on a lambda function that utilizes the SqueezeNet model for prediction. The x-axis shows a varying amount of function memory sizes and the y-axis shows the time in seconds. The figure shows the average latency in seconds as observed by the JMeter client, and the average prediction time. Figures \ref{resnet-scalable} and \ref{resnext-scalable} show similar results for the ResNet and ResNeXt models, respectively. 

\vspace{-0.1in}

The figures show that as memory increases, the latency and prediction times decrease. The platform seems to scale with demand particularly for large memory sizes where the latency is typically under an acceptable user expected response time.

\subsection{Discussions and limitations} \label{sec:discussion}
Our initial evaluation results show the feasibility of using serverless computing for prediction serving of trained neural networks. In this section we discuss our results and highlight limitations in serverless computing frameworks that need to be addressed.

The latency of warm requests is within a reasonable range. That is particularly true when the memory sizes selected are greater than 1024MB. The latency of cold requests can be significantly higher, and we assert that this need to be resolved if there we are going to couple network serving applications or machine learning applications with any form of Service Level Agreements (SLAs). While the overhead of bootstrapping a container is decreasing, it is still relatively high. One way to offset the delay is to provide access to GPU resources for prediction services because they have shown to better performance/cost ratios than CPUs. That would provide orders of magnitude improvements for serverless predictions, as reported in  \cite{mxnetperf}, where predict time is in tens to hundreds of milliseconds.

Even though the amount of resources available in serverless computing environments are increasing all the time, the resources currently available are still limited. For example, ephemeral disk capacity available for AWS Lambda functions is limited to 512MB, which limits the use of serverless platforms to serve with large neural network models, which can be larger than 500MB.

As our experiments highlighted, increased cost is not always correlated with better performance. There is a need for tools that analyze previous function executions and suggest changes in declared resources. Another option would be to scale the container vertically \cite{al2017autonomic} for optimal cost/performance based on a customer's predefined budget and performance targets. 

\section{Related work} \label{sec:relatedwork}
AWS Lambda is being used to support a wider range of applications and use cases \cite{baldini2017serverless,Chard2017WOSC,yan2016building,baldini2016cloud,bila2017leveraging}. Yan et al. design a serverless chatbot framework~\cite{yan2016building}, Baldini et al. use serverless as a backend for mobile applications~\cite{baldini2016cloud}. Chard et al. describe Ripple, a data storage architecture based on a serverless platform~\cite{Chard2017WOSC}, while Bila et al. propose a security analytics service based on OpenWhisk and Kubernetes~\cite{bila2017leveraging}. While using Virtual Machines or containers for serving models is the norm, the use of serverless for model inferencing has been previously proposed~\cite{lambdamxnet,lambdatensorflow}. An attempt to show latency is shown in \cite{lambdamxnet}, but the experiment did not isolate the specific execution times of functions and included the delay to download the model from S3. Furthermore, the experiment did not consider several dimensions to such as memory, changes in the request rate, cold and warm starts, hence are incomplete.

%However we are not aware of any performance evaluation of such a system.
% VM: Is that last statement true.

There are a number of model serving frameworks~\cite {crankshaw2017clipper,tensorflowserving,crankshaw2014missing}. Tensorflow Serving~\cite{tensorflowserving} is the open source
prediction serving system developed by Google for TensorFlow models, Velox~\cite{crankshaw2014missing} is a Berkeley research project to study personalized prediction serving with Apache Spark. Clipper~\cite{crankshaw2017clipper} is a general purpose low-latency prediction serving system. Unlike the serverless computing platform these designs were either focused on a specific deep learning framework such as Tensorflow, or were highly optimized using caching, batching, and adaptive model selection techniques. They are designed to maximize throughput and minimize latency but not necessarily to minimize operational costs when demand for the service is quickly changing or even unpredictable.

McGrath and Brenner~ propose the design of a serverless computing platform and attempt to measure performance of serverless execution across multiple cloud providers\cite{mcgrath2017serverless}. Their performance analysis focused on the execution of serverless functions of small code snippets. Our serverless function is more complicated and our experiments attempt to evaluate the feasibility of using serverless to serve deep learning models. 

\section{Conclusion and Future work} \label{sec:conclusion} 
In this work, we study the suitability of using a serverless platform for AI workloads. In particular we evaluate the performance of serving deep learning models, where a serverless function classifies images by performing a forward pass through the model. Our results indicate that warm serverless function executions are within an acceptable latency range, while cold starts add significant overhead. This bimodal latency distribution can risk the adherence to SLAs that don't take this into account. Lack of access to GPUs in serverless frameworks and the stateless nature of functions mean that each execution of the function can only use CPU resources and cannot rely on the serverless platform runtime to maintain state between invocations to optimize performance.

Our on-going research work focuses on extending our evaluations using different deep learning frameworks (e.g. Tensorflow), on different serverless computing providers and different types of models ({\it e.g.}, speech and text). Our future research work aims to better understand the trade-offs between cost and memory allocations to build customer tools to increase/decrease memory for optimized latency or cost.  

In the longer term, serverless platforms will need to support more stateful workloads such as the ones we examined. In particular, providing a declarative way to describe workloads (e.g., the minimum time to keep warm containers) and their requirements (e.g., access to GPUs) will enable performance that is close to the current state-of-the-art non-serverless platforms while still offering more flexibility around cost and scaling. A declarative approach could be extended to edge computing with serverless functions running on smartphones and IoT devices running neural compute logic~\cite{a11bionic,rpineural}.

In the meantime, using spot markets and experimenting with on-demand virtual machines with fine-grained billing, in the order of seconds, may be attractive. Eventually developers may be able to specify what AI workload they require and next-generation frameworks will run workloads as a mix of highly-optimized virtual machines with serverless filling scaling gaps, supporting both occasional prediction requests and sustained high-throughput workloads with large peaks.

% \section{appendix}
% Not sure if we should keep include this, but we also run experiments using the following step function. we can include the graphs in an extended version which we can put on arxiv.

% \begin{figure}[h]
% \centering
% \includegraphics[width=0.50\textwidth]{jmeterstepfunction2.jpg}
% \caption{jmeter second step functions}
% \label{jmeterstepfunction2}
% \end{figure}

% \begin{figure}[h]
% \centering
% \includegraphics[width=0.50\textwidth]{squeezenet-scalability2.png}
% \caption{Scalable 2 experiments squeezenet}
% \label{squeezenet-scalable2}
% \end{figure}

% \begin{figure}[h]
% \centering
% \includegraphics[width=0.50\textwidth]{resnet-scalability2.png}
% \caption{Scalable 2 experiments resnet}
% \label{resnet-scalable2}
% \end{figure}

% \begin{figure}[h]
% \centering
% \includegraphics[width=0.50\textwidth]{resnext-scalability2.png}
% \caption{Scalable 2 experiments resnext}
% \label{resnext-scalable2}
% \end{figure}

%http://docs.aws.amazon.com/lambda/latest/dg/limits.html

\bibliographystyle{IEEEtran}
\bibliography{sample-bibliography} 

\end{document}